\documentclass[10pt]{revtex4}
\usepackage{graphicx,color,comment}
\usepackage{csquotes}

\usepackage{kpfonts}

\def\t{{\cal D}}
\def\T{{\cal T}}
\def\S{{\cal S}}
\def\C{{\cal C}}

\definecolor{airforceblue}{rgb}{0.36, 0.54, 0.66}
\definecolor{babyblueeyes}{rgb}{0.63, 0.79, 0.95}

\definecolor{mediumblue}{rgb}{0.0, 0.0, 0.8}

\definecolor{amber}{rgb}{1.0, 0.75, 0.0}
\definecolor{fluorescentorange}{rgb}{1.0, 0.75, 0.0}
\definecolor{goldenpoppy}{rgb}{0.99, 0.76, 0.0}
\definecolor{process}{rgb}{1.0, 0.94, 0.0}
\definecolor{darkscarlet}{rgb}{0.34, 0.01, 0.1}

\begin{document}

\thispagestyle{empty}

\title{
{\it In Silico} Implementation of Evolutionary Paradigm in Therapy Design:\\
Towards Anti-Cancer Therapy as Darwinian Process
}

\author{B. Brutovsky}
\affiliation{Department of Biophysics, Faculty of Science, Jesenna 5,
P.~J.~Safarik University, Jesenna 5, 04154 Kosice, Slovakia}

\author{D. Horvath}
\affiliation{Technology and Innovation Park, Center of Interdisciplinary Biosciences,~P.~J.~Safarik University, 
Jesenna 5, 04154 Kosice, Slovakia}

\begin{abstract}
In here presented {\it in silico} study we suggest a way how to implement the evolutionary 
principles into anti-cancer therapy design. We hypothesize that instead of its ongoing supervised 
adaptation, the therapy may be constructed as a self-sustaining evolutionary process 
in a dynamic fitness landscape established implicitly by evolving cancer cells, microenvironment 
and the therapy itself. For these purposes, we replace a unified therapy with the `therapy species', 
which is a population of heterogeneous elementary therapies,
and propose a way how to turn the toxicity of the elementary therapy into 
its fitness in a way conforming to evolutionary causation.
As a result, not only the therapies govern the evolution of different 
cell phenotypes, but the cells' resistances govern the evolution 
of the therapies as well.
We illustrate the approach by the minimalistic {\it ad hoc} evolutionary
model. Its results indicate that the resistant cells could bias the evolution towards 
more toxic elementary therapies by inhibiting the less toxic ones.
As the evolutionary causation of cancer drug resistance has been intensively studied 
for a few decades, we refer to cancer as a special case to illustrate purely theoretical
analysis.

\end{abstract}

\maketitle

\twocolumngrid

\section{Introduction}

Development of up-to-date anti-cancer therapies relies on the deep knowledge
of specific biochemical machinery of cancer cells. However, despite
improved understanding of the molecular details of cancer initiation 
and progression, many targeted therapies fail due to diversity 
of strategies of resistance deployed by cancer cells 
\cite{Gottesman2002,Bedard2013,Gallaher2018,Gatenby2018_ColdSpring}.
Presently, intratumor heterogeneity is viewed as the principal obstacle
in the therapy design and many papers have reviewed its causes and
consequences to therapeutic resistance during the last decades
\cite{Turner2012,Saunders2012,Marusyk2010,Marusyk2012,Marusyk2013_Science,Meacham2013,Burrell2013,Fisher2013}.
The mechanisms of resistance relate to altered activity of the specific
enzyme systems, blocked apoptosis, developing transport mechanisms which 
provide multidrug resistance, etc. It was demonstrated that clonal
that accumulation of the viable clonal genetic variants poses 
greater threat of progressing to cancer than the homogenizing clonal 
expansion \cite{Maley2006}. Moreover, it
is known for a long time that the epigenetic changes, such as DNA methylation,
histone modifications, chromatin remodeling, and small RNA molecules, play
the causative role in cancer initiation, progression \cite{Laird2005,Bjornsson2004,Easwaran2014}
and resistance \cite{Hoelzel2013}.
Considering the timescales during which mutations spread in a cancer cell population,
the contribution of non-genetic instability in the intratumor heterogeneity
of cancer cell populations is significant \cite{Huang2013_CancerMetastasisRev}.

Another obstacle to efficient therapies consists in a static manner of the
administration of most therapies which underestimates that many properties of cancer 
cells that contribute to the invasion, metastases and resistance likely arise 
as successful adaptive strategies to survive and proliferate within the temporally
unstable micro-environmental conditions \cite{Gillies2018}, induced,
eventually, by the therapy itself.
It was shown that the adaptive therapeutic intervention that 
reflects the temporal and spatial variability of the tumor microenvironment
and cellular phenotype may provide substantially longer survival than
the standard high dose density strategies \cite{Gatenby2009b}.

\medskip

{\bf Evolutionary Dynamics of Cancer.}
Since Nowell conceptualized carcinogenesis as the evolutionary process
\cite{Nowell1976}, evolutionary theory has been accepted as the appropriate
conceptual base to get an insight into {\it the modus operandi} of cancer
\cite{Nowell1976,Merlo2006,Greaves2007}. Evolutionary dynamics, in which
the intratumor heterogeneity plays the crucial role, equips evolving populations
of neoplastic cells with the adaptive power enabling them to cope with
uncertain or time-varying micro-environmental conditions and it is
considered as the main reason why the targeted therapy of cancer fails
\cite{Gillies2012}, and why the combination therapy, despite often improved
therapeutic outcome, is still not the ultimate winner in the fight against
cancer \cite{Bozic2013}.
Nowadays, an effort to address the heterogeneity and variability
of cancer cells in the therapy design is apparent 
\cite{AmiroucheneAngelozzi2017,Zhang2017}.

Therapeutic resilience of advanced cancers may be attributed not only to genetic
diversity but to epigenetic plasticity as well \cite{Greaves2015_CancerDiscov}.
A major difference between the epigenetic and genetic changes is that
the epigenetic changes are reversible and, in principle, responsive to 
environmental influence.
Variability in the phenotypic characteristics of isogenic cells confers to cellular 
tissues important properties, such as the ability of cancer cells to escape 
a targeted therapy by switching to an alternative phenotype
\cite{Kemper2014,Emmons2016}.
It motivates an effort to stimulate (or prevent) the specific phenotype
switching purposefully as a therapeutic strategy
\cite{SaezAyala2013,Montenegro2015}.

The interplay between the respective genomes, epigenomes, transcriptomes and proteomes
constitutes a 'cell-state' \cite{Clevers2017}.
Due to their tendency to be self-stabilizing, there are typically fewer
distinct cell states in a tumor than it could be inferred from the degree of the genetic,
epigenetic and transcriptional heterogeneity and, straightforwardly, genetically
distinct cells may be susceptible to the treatment with the same drugs \cite{Alizadeh2015}.
On the other hand, even genetically identical cells can, due to the epigenetic differences 
and influence of an microenvironment, exist in different cell states. 
Studying the cell-state dynamics of the isogenic population of human breast cancer 
cells revealed that the three phenotypic fractions (stem, basal and luminal) stay
under the fixed genetic and environmental conditions in the equilibrium proportions and 
that individual cells transition from one state to another with constant
interconversion rates \cite{Gupta2011}. Moreover, if the fractions 
were purposely deviated from the equilibrium, the equilibrium proportions were reestablished 
by interconversions between the cell states instead of differential growth 
of the respective phenotypic subpopulations. Therefore, Markov process was
proposed as the appropriate mathematical model for the respective cell-state 
dynamics \cite{Gupta2011}.

Presuming that the cells in different states differ in their growth properties,
the cell-state composition of the cancer cell population
becomes evolutionary important trait at the cancer-relevant timescales
during which the cancer cells are exposed, in general, to changing environment
(including the therapy).
In this case, the population may benefit from maintaining 
the diversity of cell states, each advantageous within a different context
\cite{Carja2017,Dueck2016}.
It was observed that in the case of variable selective pressure, the population
of different organisms evolves the mechanisms to tune the phenotypic variability
of the population to reflect the variability of the acting selective pressure 
\cite{Rando2007}. In bacteria,
the well known risk-diversification strategy evolved in populations when
facing changing environment \cite{Crean2009,Forbes2009,Beaumont2009} is
bet-hedging \cite{DeJong2011,Donaldson2008}. This strategy increases
the long-term survival and growth of an entire lineage instead of conferring
an immediate fitness benefit to one individual \cite{Carja2017}.
Based on the formal similarity of evolving cancer cell population with
bacteria, viruses or yeast, it has been recently proposed that the
structure of intratumor heterogeneity is an evolutionary trait
which evolves towards the maximum clonal fitness at the cancer-relevant timescale
in changing (or uncertain) environment and that its structure corresponds
to the bet-hedging strategy \cite{Chisholm2016,Nichol2016,Gravenmier2018,Thomas2017}
which has been recently put into therapeutic context \cite{Mathis2017}.

\newpage

Distinguishing between the intratumor heterogeneity due to the differences
in the DNA sequences and that resulting from the epigenetic modifications is
instructive for the biological insight as well as for the 'physical'
realization of an eventual therapy.
Nevertheless, as the genetic and epigenetic changes differ primarily
in their stabilities and characteristic timescales (both can be formulated
probabilistically) and their contributions to the cell states may be 
intertwinned, the two physical levels, genetic and epigenetic, need not 
be viewed separately \cite{Alizadeh2015}.
For example, within the mathematical Markov model of the cell-state dynamics,
the probabilities of transitions can be expressed by the elements of the
transition matrix, not regarding whether genetic or epigenetic. 
Whether the epigenetic states are sufficiently stable \cite{Herman2014}
(i.~e.~whether probabilities of the respective transitions are low enough)
to enable Darwinian evolution of isogenic cells underpinned by purely
epigenetic states is an open question.

\medskip

{\bf Evolutionarily Motivated Therapies.} 
Cancer cells continuously evolve their ability to survive in time-varying
microenvironment (as exemplified by developing the resistance against eventual
therapeutic interventions). Consequently, to stay efficient, the therapy must 
be appropriately modified as well, which is typically arranged by the combination
therapies and/or appropriate scheduling schemes which make attempts to solve
this problem explicitly.
In the strategy of benign cell boosters Maley and Forrest proposed
firstly to increase the proliferation rate of the benign cells sensitive
to a cytotoxin intentionally and then apply the toxin \cite{Maley2004}.
Similarly, Chen et al. designed strategy of an 'evolutionary
trap' which selects from a karyotypically divergent population the
subpopulation with predictably drugable karyotypic feature \cite{Chen2015}.
In the evolutionary double bind strategy to control cancer, Gatenby
at al. exploit that the therapy resistance requires costly
phenotypic adaptation that reduces the fitness of the respective cells
\cite{Gatenby2009}.
It has been shown recently that the proliferation of malignant cells can
be decreased by the administration of non (or minimally) cytotoxic ersatzdroges
\cite{Kam2015,EnriquezNavas2015}
thereby the cell's resources are diverted from the proliferation and invasion 
towards the efflux pump activity, which, consequently, lowers the fraction of the cells
with developed drug efflux mechanisms in the population \cite{EnriquezNavas2015}.
During recent years, directed evolution of oncolytic viruses has been
investigated in the virotherapy \cite{Sanjuan2015}. Instead of detailed
knowledge of the molecular aspects of the interaction between the cancer
cell and the virus, the approach exploits evolutionary principles
such as diversified population of viral candidates which undergo 
purposefully designed selection steps to direct evolution towards
an explicitly pre-defined goal. Usefulness of the approach was demonstrated
by the adaptation of the RNA virus to the cells in which the
tumor suppressor gene p53 had been inactivated \cite{Garijo2014}.

Virotherapy is a targeted anticancer strategy in which genetically 
engineered strains of viruses replicate and lyse tumor cells \cite{Jenner2019}. 
Unfortunately, due to evolution of cancer cells the ultimate success of
the treatment remains elusive. We hypothesize that even if one
admits evolution of the virus, its fitness does not sufficiently reflect 
its ability to lyse cancer cells which, consequently, vanishes.
In our work, the therapy is conceived as a self-sustaining evolutionary process 
in dynamic fitness landscape \cite{Holland1975,Levins1968,Branke2002,Morrison2004}
with cytotoxicity of elementary therapies reflected in their respective fitness, 
which could prevent (or reduce) its decrease during eventual treatment.

\section{Model}
\label{model}

\begin{figure}[h!]
\includegraphics[scale=0.52]{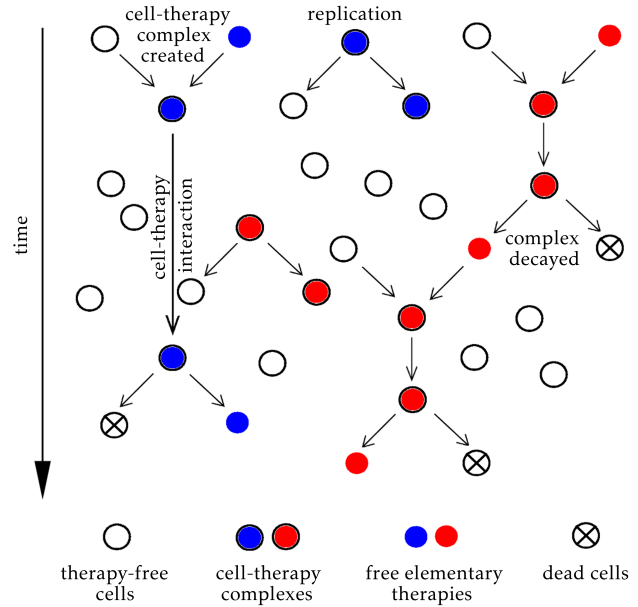}
\vspace{-5mm}
\it{
\caption
{How toxicity turns into fitness. The red elementary therapy kills the cell
sooner than the blue one therefore it can be applied more often. It
means that the toxicity plays the role of reproduction fitness.
If the therapy is not toxic, it stays within the complex and 
enables binding opportunity to free (i. e. more toxic) therapies.
}
\label{schema}
}

\vspace{-5mm}

\end{figure}

{\bf Evolutionary Causation of Toxicity.}
The ultimate goal of cancer research is to design the therapy
which stays efficient against heterogeneous and continuously changing
cancer cells with the minimum harm to healthy cells.
To implement the evolutionary paradigm into the therapy design, 
we construct the therapy as a population of heterogeneous 'elementary' 
therapies (a 'therapy species') each of them allowed to interact exclusively
with one of the available (therapy-free) cells (Fig.~\ref{schema}).
Regarding the therapeutic context, the principal feature of the elementary
therapy is its toxicity to cells. To undergo continuous, self-sustaining 
adaptation by the evolutionary process, toxicity of the elementary 
therapy must play the role of the therapy's fitness, which means that 
it must conform to evolutionary causation.

Firstly toxicity of the elementary therapy must result from the real
selection pressure, which means that the elementary therapy must be put into
interaction with the cell. Secondly the evolutionary causation requires 
that during the treatment the fundamental evolutionary principles,
phenotypic variation, differential fitness and heritability 
of fitness \cite{Lewontin1970}, apply. It follows that 
i) the elementary therapies differ in their respective toxicities,
ii) the more toxic elementary therapy is, the more often it is applied,
and, iii) repeated application of the same elementary therapy provides 
toxicity similar to that in its previous (i. e. 'parent') application,
even if it is slightly changed ('mutated'). While the points i) and iii)
are straightforward and can be ensured by the appropriate 
heterogeneity of the elementary therapies, the point ii) is less intuitive
and below we outline its minimalistic {\it in silico} 
implementation (Fig.~\ref{schema}).

In here proposed approach, each elementary therapy may create complex with 
one of the therapy-free cells. The elementary therapy
can create the new complex with another therapy-free cell only after 
its current complex has decayed, playing the role of a catalyst.
It follows that each cell-therapy complex is exposed, at the same time, 
to two counteracting selection pressures. On the one hand, the fitness 
of the cell is proportional to the number of its copies, therefore longer lifetime 
of the complex (hence the resistance to therapy) is supported by 
the evolution of the target cell population.
On the other hand, the complexes are under selection pressure due to the evolution 
of therapies; the sooner the elementary therapy kills the cells within its 
consecutive complexes, the higher is the number of its repetitions (hence,
by definition, its reproduction fitness). Consequently, shorter lifetime 
of the complex (hence higher toxicity of the elementary therapies) is 
implicitly supported by the evolution of therapies.
Despite the two evolutionary processes differ in their respective
reproduction mechanisms, the former creating physical copies (cells),
the latter repeating the therapies accordingly to their respective 
toxicities, the both processes satisfy one of the fundamental principles
of evolution, the heritability of fitness, which does not require 
any particular mechanism of inheritance; only the correlation between
the parent's and offspring's fitness is required \cite{Lewontin1970}.
To sum up, as in here proposed algorithm the number of iterations of 
an elementary therapy (hence its fitness) depends on the ability of the
therapy to kill the cell, the toxicity of the elementary therapy 
can be identified with its fitness.

To illustrate the above conceptual approach (Fig.~\ref{schema}), we present
in the following subsections  its {\it in silico} implementation.
We have devised two {\it ad hoc} models, the conceptual model of the target
cell population and the model of a virtual elementary therapy and 
its interaction with the cell within the complex. Feasibility of the respective algorithmic
steps at biochemical and biological levels is not considered; short discussion 
about an eventual implementation of the approach is provided in the Conclusions.

\medskip

{\bf Model of the Target Cell Population.}
During a few last decades many mathematical cancer models have been 
constructed, mostly classified as (i) continuum, formulated 
through differential equations, (ii) discrete lattice models, usually
represented by cellular automata and, (iii) agent based with 
diverse levels of 'intelligence' assumed in the agents 
(for complete review of mathematical cancer models, see
\cite{Anderson1998,Preziosi2003,Bellomo2008,Lowengrub2010,Wodarz2014,Altrock2015}). 
The respective types of models incorporate different levels 
of biological punctuality focusing on different scales, from microscopic
to phenomenological. As in carcinogenesis many biological phenomena run
concurrently, the multiscale models make an effort to bridge the gap between 
the phenomenology and microscopic theory \cite{Deisboeck2011,Cristini2010}. 
Evolutionary models of cancer do not typically focus on the microscales.
The reason is that the phenotypes of cancer and normal cells can have many
alternative genetic (and epigenetic) causes. Instead, evolution of cancer is driven 
by the environmental selection forces that interact with individual cellular
strategies (or phenotypes) \cite{Gillies2018}. In this way, the evolutionary
models investigate how selection pressure influence the life history
characteristics, such as survival strategy, reproduction behavior, population 
heterogeneity, etc. 
\cite{Aktipis2013_nrc,Korolev2014,Abbott2006,Maley2017}.

The {\it ad hoc} choice of the genome-coded actions in our coarse-grained 
model of the cell reflects here adopted epitomization of target cells by cancer cells.
As avoiding programmed cell death (apoptosis) is one
of the hallmarks of cancer \cite{Hanahan2011} no gene for it is assumed
in the above set of actions; target cells die due to the lack of resources
or to the therapy. On the other hand, the dormancy and phenotypic switching 
are included as they are often referred as possible alternative ways of drug
resistance.
As a result each target cell is defined by its state, $\phi\in\{0,1\}$, 
enabling to quantify the impact of an elementary therapy 
on the cell, see below, and its 'genome' consisting of $L$ genes
$g_j\in\{{\rm R,M,S,D}\},\ j=1\ldots L$, each representing specific
cell-related action. 
The tuple ${\cal G}\equiv\langle L_R/L,L_M/L,L_S/L,L_D/L\rangle$, $L_x$
being the number of gene $x$ in the genome, is introduced to express
proportions of the respective genes  in the cell's genome.
The actions associated with the respective genes correspond to:
{\bf (R)eplication:} the copy of the cell is created, unless the lack of
resources prevents it, in which case the cell
itself dies; if successful, the copy (`child') inherits the parent's
genome and state and undergoes mutation,
{\bf (M)utation:} the gene is selected uniformly at random from the genome.
If the selected gene is 'M', another gene selected uniformly at random 
is replaced by the gene picked with the same probability from $\{{\rm R,M,S,D}\}$,
{\bf (S)witching:} the cell state is switched either from $1$ to $0$ or
from $0$ to $1$, depending on the current state, and
{\bf (D)ormancy:} no action is performed.
Proportions of the respective gene in the genome
of the cell determine its phenotype such as, for example,
proliferative (abundance of 'R' genes), survival (abundance of 'D'
genes), the cell state switching ('S' gene) and genetically unstable
('M' gene). 

\medskip

{\bf Model of the Elementary Therapy.}
In our model, a unified, biochemically reasoned therapy is replaced by 
the `therapy species', which is evolving population of heterogeneous
elementary therapies. Conceptual novelty of the approach (Fig.~\ref{schema})
consists in avoiding the necessity to specify explicitly the differences
between cancer and normal cells, which is, due to intratumor heterogeneity,
the principal obstacle in common treatments. Instead, it is required 
that the model of elementary therapy introduces variability in the 
cells' survival. It follows that to prevent 'premature convergence' 
(hence genetic drift) the population of elementary therapies must start 
as less (or non)toxic.

To facilitate the instructive {\it in silico} modeling, the elementary
therapies are of the same mathematical structure based on
two characteristics - the rate of change and the selectivity.
These were chosen because they obviously influence the fate of the 
target cell as defined above, hence they guarantee the variability of
the fitness within the population of elementary therapies.  
Nevertheless, in the eventual applications much less intuitive characteristics
can be chosen, based on different pharmacological mechanisms of action.

The elementary therapy integrates all factors that effect the cell's fate 
into one time-dependent variable ${\cal E}(t)$, chosen to change 
continuously between 0 and 1 accordingly
\begin{equation}
\label{environment}
{\cal E}(t) = \frac{1 \pm \sin(t/\T)}{2},
\end{equation}
$2\pi\T$ being the period of the respective therapy.
The choice of the $+/-$ sign is arbitrary, nevertheless once chosen the sign persists
during simulation.

To quantify selectivity of the therapy, $\S$,
the threshold $\sigma$ is defined as
\begin{equation}
\label{sigmoid}
\sigma = \frac{1}{1+e^{\S(|{\cal E}(t)-\phi|-{\cal C})}},
\end{equation}
where $\C = 0.5$ postulates the symmetry of ${\cal
E}(t)$ regarding the two possible cell states.
In the next, we denote the therapy as the tuple
$\t\equiv\langle\T,\S\rangle$.

The threshold $\sigma$ determines the fate of the cell,
applying the rule
\begin{equation}
\label{action}
{\rm action} =
\left\{  
\begin{array}{l}
{\rm cell \ death}, \ \ \mbox{if} \ \upsilon_{rand}\!>\!\sigma,\\
\hspace{5.5cm},\\
{\rm selected\ uniformly\ at\ random}\\
{\rm from \ the \ genome,\  otherwise}
\end{array}
\right.
\end{equation}
where $\upsilon_{rand}\in(0,1)$ is uniformly distributed random number.
Interaction of the therapy with the cell (Eqs.~\ref{environment} to \ref{action}) 
applies only if the action selected in Eq.~(\ref{action}) is the replication.

We emphasize that in the model the cell's resistance is not bound to a
specific cell state $\phi$
but it is rather viewed as the ability of the cell(s) to survive under
an instant therapy. 
For convenience, the cell in the state $\phi$ closer to ${\cal E}(t)$
is assigned with lower probability of death (Eqs.~\ref{sigmoid},~\ref{action}).
Regarding here followed therapeutic context it is viewed as {\it resistant},
while the cell in the complementary state (farther from the ${\cal E}(t)$, 
higher probability of death) is referred as {\it sensitive}.
Taking into account the time variability of the therapy
(Eq.~\ref{environment}) it follows that the cell can be resistant in one moment
and sensitive in the other without switching its state.

\medskip

{\bf Simulation Scheme.}
The simulation begins with the population of $N$ cells characterized by their genomes 
${\cal G}_i$ and states $\phi_i,\ i=1\ldots N,$ evolving simultaneously with
the heterogeneous population of $K$ therapies $\t_k\equiv\langle\T_k,~\S_k\rangle,\ k=1\ldots K,$ 
(Eqs.~\ref{environment},~\ref{sigmoid}).
Each genome ${\cal G}_i,\ i=1\ldots N,$ consists of 
$L$ genes $g_j,\ j=1\ldots L,$ selected, at the beginning, 
uniformly at random from $\{{\rm R,M}\}$. The states 
$\phi_i, \ i=1\ldots N$ are picked from $\{0,1\}$ with the same probability.
The therapy species is represented by $K$ elementary therapies 
$\t_k = \langle\T_k,\S_k\rangle,\ k=1\ldots K,$
each of them with its own period $2\pi\T_k$ and selectivity $\S_k$, starting as
\begin{equation}
\label{initpar}
\T_k = 10^{\xi}\ \ {\rm and}\ \ \S_k = 10^{\eta},\ \ \ \ \ k=1\ldots K,
\end{equation}
where $\xi\in(\xi_{min},\xi_{max})$ and $\eta\in(\eta_{min},\eta_{max})$
are uniformly distributed random numbers.

During simulation, the cells and therapies evolve accordingly to  
the model (Eqs.~\ref{environment}-\ref{action}).
The number of cells varies as implied by their interaction with elementary
therapies and the population carrying capacity provided by the system resources.
At replication, the newly born cell inherits its state and genome from its parent,
and the both genomes undergo the above described mutation procedure.
In addition, the offspring cell creates exclusive
lifetime complex with one of the free  elementary therapies. 
In contrast to the variable size of the population of cells, the size of the
therapy species is kept constant; elementary therapies neither replicate,
nor die; they only mutate every time when they form the complex
with the cell; the mutated therapies are obtained as
\begin{equation}
\label{mutatedpar}
\T_k^{new} = \T_k^{old}\times10^{\delta_1}\ \ {\rm and}\ \ \S_k^{new} = {\cal
S}_k^{old}\times10^{\delta_2},
\end{equation}
where $\delta_1, \delta_2\in(0,0.1)$ are uniformly distributed random
numbers.
Subsequently, the values of parameters within required intervals 
are guaranteed by imposing the limits as
\begin{equation}
\label{boundaries}
\tilde{X}=
\begin{cases} 
X\times10^{\gamma_{max}-\gamma_{min}},&{\rm for}\ \ \ X<10^{\gamma_{min}},\\
X\times10^{\gamma_{min}-\gamma_{max}},&{\rm for}\ \ \ X>10^{\gamma_{max}},\\
X,&{\rm otherwise}
\end{cases}
\end{equation}
where $X$ stands for $\T_k^{new}$ or $\S_k^{new}$ and
$\gamma$ for the respective $\xi$ or $\eta$ in Eq.~\ref{initpar}.

During its lifetime, the cell repeatedly chooses uniformly at random
the gene from its genome and performs the respective actions.
If the action is replication,
the interaction of the cell and its respective elementary therapy 
is recalculated (Eqs.~\ref{environment}-~\ref{action}).
If the cell survives, it replicates.
When the cell dies, either due to the interaction with the therapy or due
to the lack of resources at the moment of replication, its respective 
complex decays and the relinquished elementary therapy becomes capable 
to create complex with another newly born cell, playing the role of a catalyst.

\section{Results}
\label{results}

Here, the cell state dynamics is represented by the dependence of the ratio $N^1/N$
on the variable ${\cal E}(t)$; $N^1$ and $N$ are numbers of the cells
in the state $\phi = 1$ and the target cell population size, respectively.
Due to fundamentally different physical implementation of the model system,
with biological time scales and carrying capacity substituted by the CPU and
memory limits (see Appendix), numerical values of $\T$ and $\S$ in the figures
below lack biological meaning and are used only to outline a few typical 
behaviors of the model.
Despite that, some results, such as the dependence of the cell state 
dynamics on the relations between some of the parameters could have 
universal meaning.
In the subsections C and D, ${\cal E}(t)$ (Eq.~\ref{environment}) was 
calculated for each cell-therapy complex separately, with time $t$ corresponding 
to the cell's age expressed as the CPU time consumed by the cell's thread 
(running in parallel with other threads, see Appendix) instead
of simulation (i. e. 'physical') time, providing ${\cal E}(0) = 0.5$ at the
cell's birth.
Consequently, the ranges of $\T$ of elementary therapies chosen in the 
Results sections A and B differ in orders of magnitude from those 
in sections C and D.

To get deeper insight into below {\it in silico} investigation of the evolutionary
dynamics of the cells and elementary therapies, we have firstly studied 
a few cases of the cell state dynamics where at least one population, cells or
therapies, was kept homogeneous (i. e. isogenic or unified, respectively) and did not evolve.
Though most of presented features are rather obvious by intuition,
for the readers' convenience we point out the most dominant of them.

\begin{center}
{\it A. Isogenic target cell population under unified therapy}
\end{center}

\vspace{-1.2mm}

\begin{figure}[h!]
\hspace*{-5mm}\includegraphics[scale=0.83]{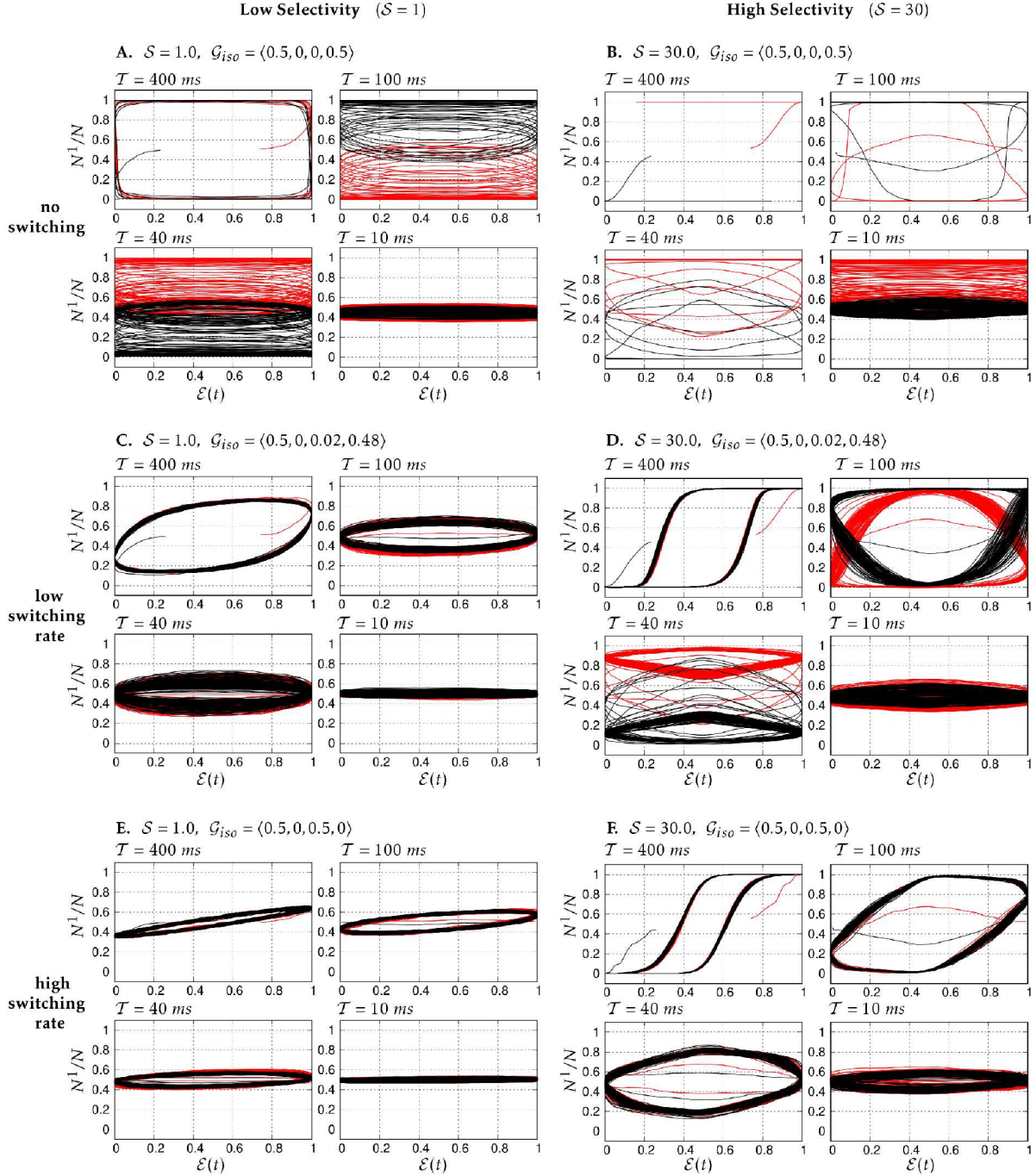}
\vspace{-3mm}
\it{
\caption
{ 
Examples of the cell state dynamics of non-evolving populations of isogenic
cells which differ by the probability of the cell state switching - no switching, low
($L_S/L=0.02$) and high ($L_S/L=0.5$) switching probability; 
for each of the populations two kinds of therapies are
applied - low and highly selective, respectively, quantified by the values of the parameter 
$\S = 1$ and $30$, respectively. Moreover, for all the cases four unified therapies
differing in their respective periods are applied, as quantified by the values
of the parameter $\T = 400,100,40$ and $10\ ms$ in (\ref{environment}). 
In each plot two series corresponding to the plus (red) 
or minus (black) sign in Eq.~(\ref{environment}) are depicted.
}
\label{cell_state_dynamics}
}
\end{figure}

Firstly dependences of the cell state dynamics on the three model 
characteristics, selectivity and period of the unified therapy
and the cell's switching rate ($L_S/L$ in ${\cal G}_{iso}$), were investigated.
Fig.~\ref{cell_state_dynamics} shows convergence of the $(N^1/N,{\cal E})$
trajectories for the respective combinations of the above three model 
characteristics. The three nonevolving isogenic target cell populations 
which differ each other in the cells' switching rates,
i) no switching, ${\cal G}_{iso}=\langle0.5, 0, 0, 0.5\rangle$,
ii) low switching, ${\cal G}_{iso}=\langle0.5, 0, 0.02, 0.48\rangle$, and
iii) high switching, ${\cal G}_{iso}=\langle0.5, 0, 0.5, 0\rangle$,
were exposed one by one to 8 different unified therapies 
$\t_{uni}=\langle\T,\S\rangle$, combining a few different periods $2\pi\T$
($\T = 400, 100, 40, 10\ ms$) and selectivities ($\S = 1, 30$).
We note that to satisfy the genome's constraint $(L_R+L_M+L_S+L_D)/L = 1$, the change
in the switching rate $L_S/L$ is compensated by the respective change of
the probability of dormancy, $L_D/L$. Nevertheless, taking into account 
minor role of the dormancy within our illustrative model, we omit analysis 
of this step.
Isogenic populations consisted of a few thousands non-evolving
cells in the states $\phi_i,\ i=1\ldots N,$ each picked from $\{0,1\}$
with the same probability.

\onecolumngrid

\twocolumngrid

\begin{figure}[h!]
\hspace*{-2mm}
\includegraphics[scale=0.98]{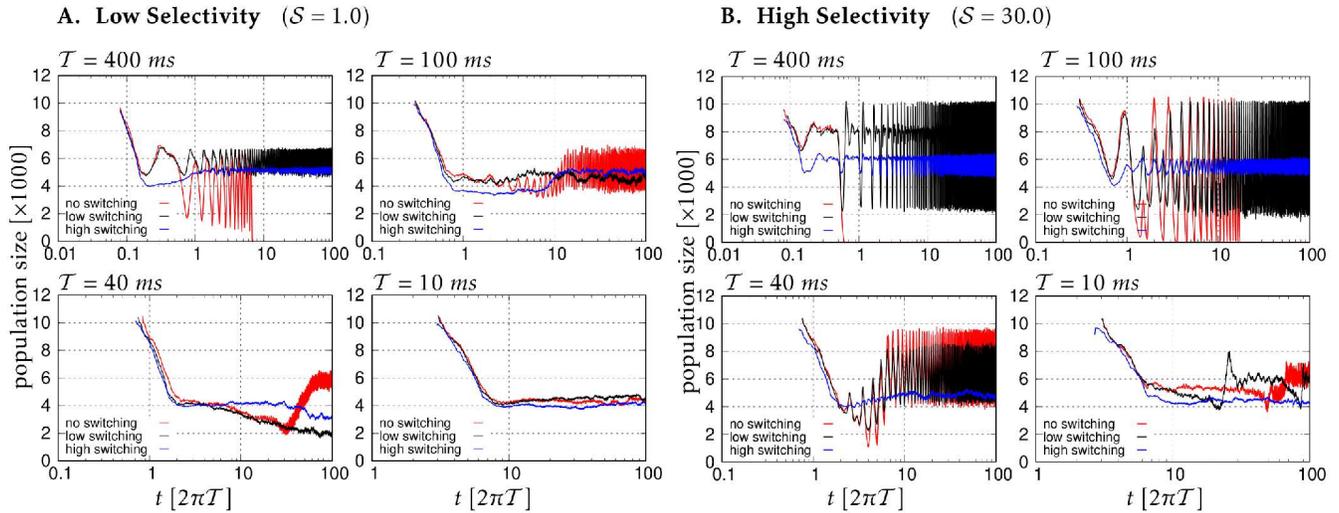}
\vspace{-5mm}
\it{
\caption
{ 
Dependence of the target cell population size on the period and selectivity of the
therapy and the rate of switching of isogenic cells. Obviously, under higher selectivity
the system follows dynamics of ${\cal E}(t)$ unless too short 
period (here, corresponding to the case $\T$ = 10 ms).}
\label{popsize}
}

\end{figure}

In the case without switching, ${\cal G}_{iso}=\langle0.5, 0, 0, 0.5\rangle$, 
the population becomes extinct if the period $2\pi\T$
is too long ($\T = 400\ ms$ in Figs.~\ref{cell_state_dynamics}A,
~\ref{cell_state_dynamics}B, ~\ref{popsize}A, and ~\ref{popsize}B) 
for the survival of the sensitive cells. 
The probability of extinction increases with higher selectivity of the therapy 
(Figs.~\ref{popsize}A and ~\ref{popsize}B).
In the case of low selectivity, $\S=1.0$, and the shortest investigated 
period given by $\T = 10\ ms$, the ability to switch the cell state does not affect 
the cell state dynamics  significantly (Fig.~\ref{cell_state_dynamics}A to C).
However, high selectivity, $\S = 30.0$, can homogenize the cell states in
the population even during very short period without switching
(Fig.~\ref{cell_state_dynamics}B, $\T = 10\ ms$).
Between the two limiting periods, corresponding to $\T = 400\ ms$ and $\T = 10\ ms$,
respectively, the population does not become extinct and converges to one of the 
two cell states.
When the phenotype switching is allowed, the typical hysteretic behavior of the cell state
dynamics emerges for the selectivity $\S=30$ (Figs.~\ref{cell_state_dynamics}D and 
\ref{cell_state_dynamics}F, the period given by $\T=400ms$) with the width of 
hysteretic loop decreasing with the switching probability.
It is obvious that the similar cell state dynamics can be produced alternatively
(Figs.~\ref{cell_state_dynamics}C,~$\T=400\ ms$ and ~\ref{cell_state_dynamics}F,
$\T=100\ ms$), which indicates dependence 
of the cell state dynamics on a scaling form constructed between
selectivity, period and the rate of switching.

We note that the relation between hysteresis and phenotype switching in
evolutionary systems has often been observed and studied
\cite{Friedman2014}. It was shown in bacteria that some antibiotics can induce
long-lasting changes in their physiology, termed cellular hysteresis,
that influence bactericidal activity of other antibiotics and can be 
exploited to optimize antibiotic therapy \cite{Roemhild2018}.
However, keeping in mind {\it ad hoc} choice of the model aimed in particular
to provide formal fitness landscapes for here investigated purposes, we leave deeper
analysis of the specific hysteretic behavior (or the memory effects caused by the
therapy) of the cell state dynamics and, eventually, the scaling properties
to future research.

\onecolumngrid

\vspace{5mm}

\vspace{2mm}

\twocolumngrid

\begin{center}
{\it B. Evolving target cell population under unified therapy} 
\end{center}

\begin{figure}[h!]
\hspace*{-3mm}\includegraphics[scale=0.91]{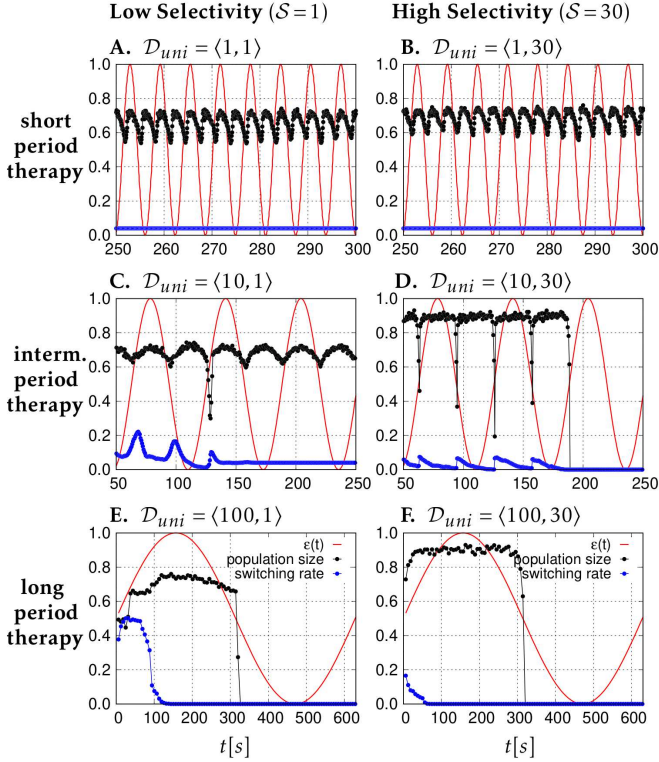}
\vspace{-5mm}
\it{
\caption
{ 
Evolution of the switching rate under 6 different unified therapies
$\t_{uni}=\langle\T,\S\rangle$, where the short period therapy uses $\T=1$s,
intermediate $\T=10$s, and the long period therapy $\T=100$s.
The sizes of the target cell population and the switching rates in
the genome after the population becomes isogenic are rescaled to emphasize 
dependences of the respective series on the variable ${\cal E}(t)$.
Population size counts the both cell states, $0$ and $1$;
however, at the ${\cal E}(t)$ extrema one of the states (alternately) dominates.
}
\label{evolution_of_switching}
}
\end{figure}

In this subsection, the target cell population (for its initialization see
simulation scheme in Section~\ref{model}) evolves under
a few unified therapies $\t_{uni} = \langle\T,\S\rangle$ which differ by 
their periods $2\pi\T$ and selectivities $\S$. 
In the case of short periods, here corresponding to $\t_{uni} = \langle1,1\rangle$ and 
$\langle1,30\rangle$, the switching probabilities $L_S/L$ in the population
converged to $0.02$ (Figs.~\ref{evolution_of_switching}A,B), which is the lowest 
possible nonzero value within the model resolution.
In the case of long periods, corresponding to the unified therapies 
$\t_{uni} = \langle100,1\rangle$ and $\langle100,30\rangle$, no switching
has evolved (Figs.~\ref{evolution_of_switching}E to F).
When the therapies with intermediate periods, $\t_{uni} = \langle10,1\rangle$ 
and $\langle10,30\rangle$, are applied, higher selectivity makes population
more state homogeneous and, at the same time, decreases switching 
probabilities (Figs.~\ref{evolution_of_switching}C, D).

In the previous subsection we demonstrated sensitivity of the cell 
state dynamics on the period $2\pi\T$ and the selectivity $\S$ of the therapy as 
implemented in the model (Eqs.~\ref{environment}-\ref{boundaries})
as well as on the switching probability.
In this subsection the interplay between the period $2\pi\T$ and selectivity 
$\S$ of the therapy and the evolution of switching probability was shown
in more depth and the consequences of evolving it (or not) for the cell state 
dynamics.
 
The results presented in this and the previous subsections are very general
and reflect simplicity of the above {\it ad hoc} models. To obtain biologically
more relevant outcome, biologically realistic timescales for the replication,
phenotypic switching, apoptosis, carrying capacity, etc, would be needed.
Nevertheless, regarding the context of our work we require the 
capability of the model to generate sufficient variability of 
the cell state dynamics, which has been demonstrated.

\begin{center}
{\it C. Isogenic target cell population under evolving therapy} 
\end{center}

Here, the relation between toxicities of the elementary therapies and their fitness 
was investigated. Toxicity of the  elementary therapy is represented by
the average lifetime of the cells which have applied it, and the fitness 
of the therapy corresponds to the number of its iterations,
i. e. its abundance within the space of all possible elementary therapies
spanned by the intervals $(\xi_{min},\xi_{max})\times(\eta_{min},\eta_{max})$
(Eq.~\ref{initpar}).

In simulations, the therapy species consisting of $K = 32768$ elementary therapies
$\t_k = \langle\T_k,\S_k\rangle, \ k=1\ldots K$, evolves in interaction with 
non-evolving isogenic populations (4 populations with different switching probabilities 
were one by one tested, ${\cal G}_{iso}= \langle0.5, 0, 0, 0.5\rangle$, $\langle0.5, 0, 0.02, 0.48\rangle$, 
$\langle0.5, 0, 0.2, 0.3\rangle$ and $\langle0.5, 0, 0.5, 0\rangle$).
At the beginning, the cell states $\phi_i, \ i=1\ldots N$, were picked 
from $\{0,1\}$ with the
same probability and the elementary therapies' parameters 
$\T_k$ and $\S_k,\ k=1\ldots K$, determining the periods and selectivities
of the respective elementary therapies (Eq.~\ref{sigmoid}), were generated accordingly to Eq.~\ref{initpar} 
for $\xi\in(2,9)$ and $\eta\in(0,3)$. 
The new ('mutated') elementary therapies were obtained from Eq.~\ref{mutatedpar}
imposing the boundary conditions (Eq.~\ref{boundaries})
with $\delta_1, \delta_2\in(0,0.1)$.

\begin{figure}[h!]
\begin{flushleft}
\hspace*{0mm}\includegraphics[scale=0.5]{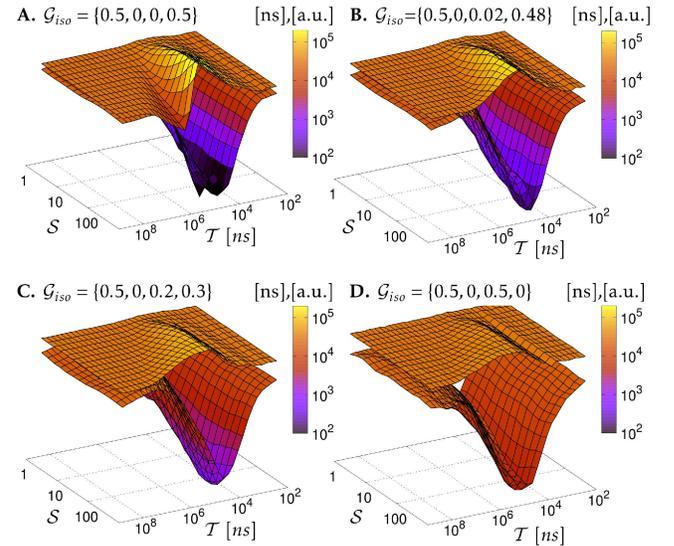}
\vspace{-5mm}
\it{
\caption
{The average lifetimes of the cells (upper surfaces) which apply the same elementary 
therapy versus number of its applications (expressed as the density of
the discretized $(\xi_{min},\xi_{max})\times(\eta_{min},\eta_{max})$
therapy space, bottom surfaces, a. u. standing for arbitrary units).
The plots A to D show the results for 4 isogenic target populations ${\cal G}_{iso}$
which differ in switching probabilities.
The surfaces corresponding to the density of therapy space are rescaled for
the demonstration purposes; the heat maps of the respective densities of the therapy space 
(bottom surfaces) are shown more precisely in
Fig.~\ref{isopop_evother_heatmaps}A to D.}
\label{isopop_evother_surfaces}
}
\end{flushleft}
\end{figure}

\begin{figure}[h!]
\begin{flushleft}
\vspace{-3mm}
\hspace*{-2mm}\includegraphics[scale=0.85]{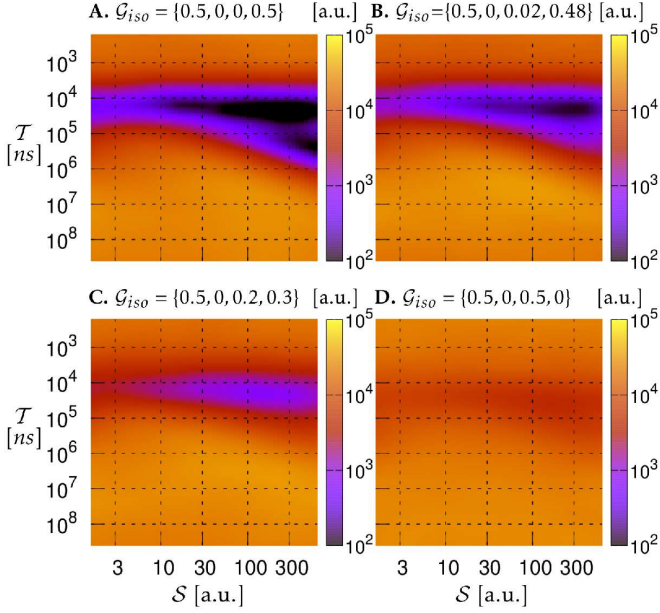}
\vspace{-5mm}
\it{
\caption
{
The heat maps show more precisely the density of the therapy space for the corresponding populations 
in Fig.~\ref{isopop_evother_surfaces}. Colors of the heat maps are scaled 
to sharpen positions of the extrema in the surfaces.
}
\label{isopop_evother_heatmaps}
}
\end{flushleft}
\end{figure}

The simulation results show that increase of the switching
probabilities in isogenic target cell populations makes the distributions
of average lifetimes, as well as the density of the therapy space,
flatter (Fig.~\ref{isopop_evother_surfaces}).
The average lifetimes of the cells under evolving therapies
presented in Figs.~\ref{isopop_evother_surfaces} and \ref{isopop_evother_heatmaps}
were $\approx 52\mu s$ for
${\cal G}_{iso} = \langle0.5, 0, 0, 0.5\rangle$, $\approx 52\mu s$ for ${\cal G}_{iso}= \langle0.5, 0, 0.02,
0.48\rangle$, $\approx 46\mu s$ for ${\cal G}_{iso}=\langle0.5, 0, 0.2, 0.3\rangle$ and 
$\approx 39\mu s$ for ${\cal G}_{iso}=\langle0.5, 0, 0.5, 0\rangle$.

The main result of this subsection (Figs.~\ref{isopop_evother_surfaces},~\ref{isopop_evother_heatmaps}) is that less toxic elementary therapies (conferring to 
cells longer average lifetimes) become less abundant in the therapy species.
Moreover, the flatness of the distribution and the average lifetime of the cells 
depend on the switching probability encoded in the genomes.

\newpage

\begin{center}
\mbox{{\it D. Evolving target cell population under evolving therapy}}
\end{center}

\begin{figure}[h!]
\hspace*{-2mm}\includegraphics[scale=0.75]{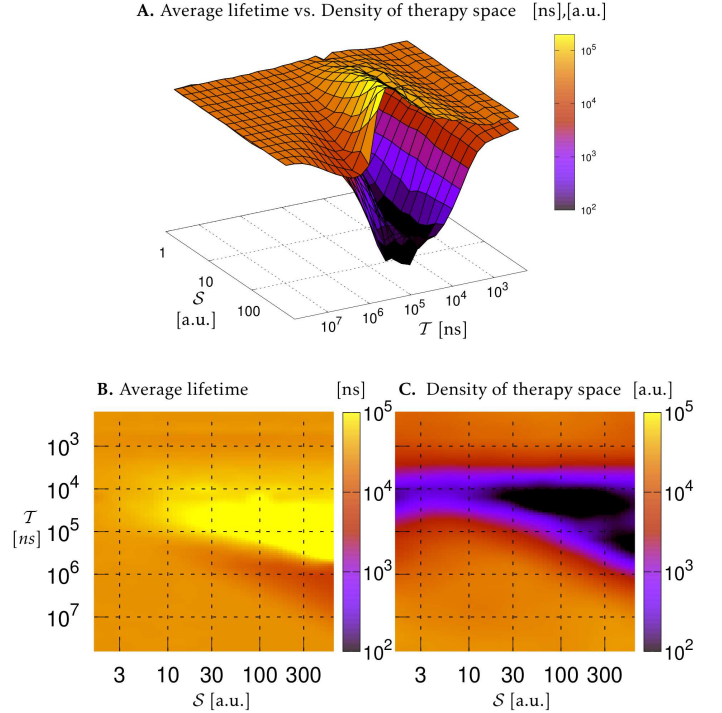}
\vspace{-5mm}
\it{
\caption
{
Comparison of the average lifetimes of the cells which applied
the same elementary therapy (upper surface in plot A) with the density 
of the space of elementary therapies (bottom surface in A, a.u. standing 
for the arbitrary units). As the density of the therapy space is recorded after the population 
has converged to isogenic (${\cal G}_{iso}=\langle0.6, 0, 0, 0.4\rangle$),
the average lifetimes depend only on the respective elementary therapies.
The corresponding heat maps B, C underline the correlation of the area
of the above-average
lifetimes (bright yellow area in plots A, B) with sparsely populated areas in the 
therapy space (deep violet area in plots A, C).
}
\label{surfaces}
}
\end{figure}

Here, the population of therapies evolves simultaneously with the population
of cells. Dependence of the average lifetimes of the cells on applied therapies,
specified by the parameters $\T$ and $\S$, was investigated. 
As the density of the therapy space is recorded after the
population of target cells has converged to isogenic
(${\cal G}_{iso}=\langle0.6, 0, 0, 0.4\rangle$)
the differences in the cells' 
lifetimes are attributed exclusively to the applied therapies.
Evolved switching probability was 0, which is consistent with the
findings in the previous subsection that, in this specific case,
the absence of switching increases the average lifetime of cells.    
Fig.~\ref{surfaces} shows that the therapies conferring, on average, longer
lifetimes to the respective cells are underrepresented in the therapy species
and vice versa.

The explanation of the anticorrelation is straightforward. Each cell,
at its birth, creates the exclusive complex with one of the available 
elementary therapies (which 'mutates').
Even neither in the case of long living cell (dormant cell 
or the cell resistant to its respective therapy) is the therapy replaced.
It might seem (therapeutically) contraproductive, as it contradicts
to an intuitive expectation that the resistant cells
should be the primary target of therapies, as resisting the cell death
is one of the hallmarks of cancer cells \cite{Hanahan2011}.
As it was discussed above, here the fitness of the therapy 
corresponds to the number of its iterations (instead of the number of its
physical copies). Straightforwardly, the therapeutic effort is to arrange that
more fit (toxic) elementary therapies are applied more often, and vice versa, 
if the therapy is not efficient, it should be repeated only rarely
(to preserve the exploration capability of the algorithm).
The anticorrelation in Fig.~{\ref{surfaces}} implies that the less
efficient therapy is, the less often it is used, which is the desired result.
The message is that if it is not possible to determine {\it a priori}
which therapy is the best (and to apply it), one should eliminate less 
fit therapies (to increase the abundance of the more toxic ones).

\vspace{-7mm}

\section{Discussion}
\label{discussion}

In here presented scenario, the resistant cells implicitly serve as
inhibitors for non (or less) efficient therapies. By this way, the resistant cells  
direct the evolution of the therapy species to more efficient therapies. 
Here, it is required that the elementary therapy can create complex with
the next therapy-free cell only after its current cell's death.
By this, toxicity of the therapy is connected with its fitness 
in the evolution conforming way.
However, this requirement disregards that many cancers evolve multidrug
resistance by upregulating membrane efflux pump that exports drugs,
thereby ensuring the cell's survival. 
As in this case the therapy is pumped out before it can fully 
exhibit its particular properties (primarily its toxicity), 
evolution of elementary therapies becomes questionable 
as one of the crucial evolutionary principle, the differential 
fitness \cite{Lewontin1970}, is significantly suppressed.
On the other hand, the efflux pump comes with the energetic cost
\cite{Silva2012} which makes the cells with the efflux pump 
less fit than those without it when the therapy is absent
(or nontoxic \cite{Kam2015,EnriquezNavas2015}).

In here presented conceptualization, the therapies are heterogeneous,
each of them interacting with the cells with heterogeneous
properties, including the differences in their sensitivity to
different therapies.
If the cell pumps out the therapies not regarding their respective
toxicities, it wastes resources and becomes less fit.
Therefore, we speculate that in reality the cells would evolve 
mechanism(s) enabling them to extrude therapies 
reflecting the level of their respective toxicity.
If true, more toxic therapies prevail in population,
just as in here investigated case when the death of the cell
was required to re-apply its therapy by the next cell.
It would mean that the toxicity plays, from the evolutionary viewpoint,
the role of the fitness of the therapy
no matter whether it shortens the cell's lifetime, or redirects the cell's resources
from replication and invasion to building efflux mechanism(s). 

In here presented algorithm, each newly born cell
creates the complex with the therapy which is, in general, different from the therapy of 
its parent. It follows, that even if the offspring has inherited resistance 
against the therapy of its parent, it might still be sensitive to its own,
in addition mutated, therapy, which obviously decreases its (as well
as its parent's) fitness. This effect might be more pronounced in the
case of cells with unlimited replicative potential, which is one of
the cancer hallmarks \cite{Hanahan2000}.
This adds more biological flavor to the model, as most chemotherapeutic 
drugs are designed to effectively target fast-dividing cells.

\section{Conclusions}
\label{conclusion}

In the paper, a unified therapy is substituted by the therapy species,
the evolving constant size population of heterogeneous elementary 
therapies, each of them with the fitness resulting {\it a posteriori}
from its interaction with the cells and vice versa.
In this way, not only the therapies govern the evolution of different
phenotypes, but the variable resistances of the cells govern the evolution 
of the therapies as well.
From the viewpoint of evolving elementary therapies, evolving population
of target cells plays the role of dynamical environment.
As the therapies themselves mutate,
less dense areas of the therapy space are repeatedly repopulated 
by ‘diffusion’, retaining exploratory power of algorithm.
Our {\it in silico} investigation indicates that the algorithm
can identify the most efficient therapies by inhibiting those which
are less efficient (as evidenced by their lower ability to kill 
the host cell). Not being tailored to some specific
molecular mechanisms responsible for the respective cancerous
features, the approach could, in principle, to cope with intratumor heterogeneity
and stay efficient during adaptation of cancer cells to changed therapy.

Despite conceptual simplicity of the above approach we foresee a number
of technical difficulties in its eventual therapeutic implementation.
The ultimate question,~i.~e.~which species will be the winner of the
evolutionary 'arms race' - cancer or therapy - stays therefore unanswered in our
paper. Each evolutionary process samples its respective search space with 
some specific efficiency.
While the efficiency of the sampling by cancer cells
population results from their biochemistry, in the case of the
therapy species the efficiency of sampling would derive from 
its eventual therapeutic realization.
The questions follow: What agent could be used as the replication-deficient
therapy species? Must it be organic at all? How to mutate the therapies?
How to deliver the elementary therapy into the cell, and, subsequently, to avoid
the efflux pump, etc.     
Some of the above issues are omnipresent in cancer research and are
intensively studied. Important
insights could be gained from the virotherapy where the evolutionary principles
are used to direct evolution towards the explicitly pre-defined goal, and the
virus-based gene-therapy which uses the replication-deficient viruses as 
vectors.

Recently, the framework for classifying tumors according to their 
evolvability was presented \cite{Maley2017}, based on the diversity of
neoplastic cells and its changes over time (constituting the Evo-index)
on the one hand, and the environmental hazard and availability of resources
(Eco-index) on the other hand. The authors of the above paper 
envision that in the future, the classification could indicate how the tumor 
of the respective type of evolvability would change
with different therapy, helping so clinicians to choose the interventions
regarding evolvability of the respective neoplasms.
Our work enables to study the interplay between related characteristics,
such as the phenotype switching and mutability, which reflect diversity 
of the target cell population and its changes over time and could eventually
play the role of the Evo-index.
Similarly, the environmental hazard is in our work represented 
by the selectivity of the therapy. The resources are implicitly 
(and, in principle, unavoidably) included in the model.
Variability of the selection pressure is mediated by the different periods
of elementary therapies.
Owing to this we believe that here presented approach could contribute 
to better understanding of the relation between evolvability of cancer
and dynamics of environment (hence the therapy). 

In his iconoclastic paper \cite{Adleman1994}, Leonard Adleman proposed
that computationally hard problems, such as therein presented NP-complete directed
Hamiltonian path problem, can be efficiently solved with the algorithmic steps 
realized by the standard tools of molecular biology. 
We hope that here speculated possible benefits of yet conceptual approach
could, perhaps, motivate cancer researchers to test its feasibility 
in a therapeutically relevant way.

\bigskip

\section*{ACKNOWLEDGMENTS}

This work was supported by the Scientific Grant Agency of the Ministry of Education
of Slovak Republic (VEGA) under the grants No. 1/0250/18 and 1/0156/18. 

\onecolumngrid

\section*{APPENDIX}

\twocolumngrid

The program encoding the above model conforms to the POSIX Threads execution
scheme with the round-robin time sharing as implemented in the Linux programming
interface API \cite{Kerrisk2010}, enabling to execute the respective cells
concurrently as separate threads.
More important than an eventual CPU gain, accruing from the program
parallelism, is the benefit from delegating some of the model's parameters to
the respective system CPU and memory resources.
For example, the size of the cell population derives, apart from the
cells' genomes themselves, from the maximum number of concurrently running
threads allowed by the system (representing `carrying capacity' of the population), 
duration of the thread creation (being the counterpart of the cell replication),
size of the thread stack, etc.
Owing to the implicit substitution of some model parameters by the system
parameters, the implementation of the model is more robust, simpler and
enable to concentrate on particular biologically relevant aspects, such as
demonstrated by the above results. Obviously, the hardware constraints
can be viewed as the counterpart of the constraints implied by biochemistry which are always
present in biological experiments.

Our specific hardware restrictions, such as the maximum allowed concurrently
running threads in the system, duration of the thread creation
(hundreds of nanoseconds) and cancellation processes, etc, enabled us to 
simulate populations close to maximum size $N \approx 10000$, Fig.~\ref{popsize}.
In summary, between $10^7$ and $10^{8}$ ($\approx 10^{4}$ per second)
cell-therapy complexes could be tested in a 1-hour simulation.
The number of elementary therapies in the therapy species during simulations was kept constant
($K = 32768)$, reflecting the particular hardware implementation.

\onecolumngrid

\end{document}